\documentclass[iopart,jpa]{iopart}
\usepackage{iopams}
\usepackage[varg]{txfonts}
\usepackage{shortvrb,psfrag,graphicx}
\def\be{\begin{equation}}
\def\ee{\end{equation}}
\def\bea{\begin{eqnarray}}
\def\eea{\end{eqnarray}}

\def\a{\alpha}
\def\b{\beta}

\def\e{\epsilon}

\def\g{\gamma}

\def\l{\lambda}
\def\m{\mu}

\def\r{\rho}
\def\s{\sigma}

\def\o{\omega}

\def\G{\Gamma}

\begin{document}
\title{Nonlinear compression of autonomous similaritons in the cubic-quintic
 nonlinear Schr\"odinger equation model with nonlinear gain }
\author{Thokala Soloman Raju}
\address{Department of Physics, Karunya University, Coimbatore 641 114, India}
\ead{soloman@karunya.edu,~solomonr\_thokala@yahoo.com}

\begin{abstract}
We describe similariton pulse propagation in double-doped optical
fibers with the aid of self-similarity analysis of the
cubic-quintic nonlinear Schr\"odinger equation with varying
dispersion, nonlinearity, gain or absorption, and nonlinear gain.
Exact similariton pulses that can propagate self similarly subject
to simple scaling rules of this model have been found using a
fractional transform. By appropriately tailoring the dispersion
profile and nonlinearity, the condition for optimal pulse
compression has been obtained. Also, nonlinear chirping of the
trigonometric solution has been demonstrated.
\end{abstract}

\maketitle

\section{Introduction}
Over three decades ago \cite{hasegawa}, a major contribution to
fiber-optics research was made when Hasegawa and Tappert pointed
out that solitons could propagate in glass fibers with anomalous
group velocity dispersion (GVD) under the influence of  the
optical Kerr effect. Optical solitons were observed in 1980
\cite{stolen}, and are of fundamental interest because of advances
in the control and generation of ultrashort optical pulses
\cite{govind1} and their potential applications in the field of
optical fiber communications. Recently, solitary wave pulses
having linear chirp and parabolic intensity profile have been
predicted and experimentally observed in an optical fiber
\cite{anderson} and in optical fiber amplifiers
\cite{Fermann,kruglov1} with normal GVD. These optical pulses
propagating in a self-similar regime were called similariton
pulses, or more pertinently parabolic similariton pulses (PSPs).
These pulses have recently found increasing practical application
in high-power amplifier systems
\cite{limpert,malinowski,peacock,finot}, efficient temporal
compressors \cite{Fermann,limpert,malinowski}, and similariton
lasers \cite{Ilday}. Another type of solitary wave pulse
propagating in self-similar regimes with linear chirp and
sech-amplitude profile has been predicted in optical fiber
amplifier with anomalous GVD
\cite{moores,serkin,kruglov,chen,chen1}. These pulses, for the
same reason were called similariton pulses, and to distinguish
them from PSPs we will also call them sech-similariton pulses
(SSPs). These SSPs propagating in self-similar regimes are of
considerable interest for pulse compression applications that
recently were demonstrated experimentally \cite{mechin}.

By now, it has been well established that cubic-quintic nonlinear
Schr\"odinger equation (CQNLSE) with nonlinearity management
presents practical interest since it appears in many branches of
physics such as nonlinear optics and Bose-Einstein condensate
(BEC). In nonlinear optics it describes the propagation of pulses
in double-doped optical fibers \cite{angelis}. In BEC it models
the condensate with two and three body interactions
\cite{abdullaev,xhang}. In optical fibers periodic variation of
the nonlinearity can be achieved by varying the type of dopants
along the fiber. In BEC the variation of the atomic scattering
length by the Feshbach resonance technique leads to the
oscillations of the mean field cubic nonlinearity \cite{inouye}.
Serkin et al described optical organic materials and waveguides
modeled by CQNLSE as the key elements for future
telecommunications and photonic technologies \cite{serkin2}. For
example thin films of polydiacetylene para toluene sulfonate
exhibit cubic-quintic nonlinearity \cite{serkin2}.  Modulational
instability of the CQNLSE through the variational approach has
been discussed \cite{ndzana}. Recently, solitons with cubic and
quintic nonlinearities modulated in space and time has been
discussed \cite{avelar}. An analytical approach to soliton of the
saturable nonlinear Schr\"odinger equation determination and
consideration of stability of solitary solutions of CQNLSE has
been discussed \cite{adib}. Also, solitary wave solutions for high
dispersive CQNLSE has been discussed \cite{triki}. Although we
discuss self-similar autonomous solutions in this model,
interested reader is referred to excellent reviews by Serkin et al
in Refs.\cite{serkin3,serkin4} on nonautonomous solitons. In this
paper, we describe a different chirped similariton pulses for the
generalized CQNLSE with varying dispersion, nonlinearity, gain or
absorption, and nonlinear gain using a recently developed
fractional transform (FT) \cite{panigrahi}. Unlike the PSPs and
SSPs, the similariton pulses found for this model are
Lorentzian-type ones some being singular, and others being
nonsingular, and periodic. Furthermore, we delineate the
compression of these exact chirped similariton pulses by piloting,
precisely, the pulse position through appropriate tailoring of the
dispersion profile. The results reported here have potential
application to the design of double-doped fiber optic amplifiers,
optical pulse compressors, and solitary wave based communication
links using higher order nonlinearity and nonlinear gain.
\section{SIMILARITON PROPAGATION IN DOUBLE-DOPED FIBERS}

As detailed in Ref. \cite{angelis}, the double-doped semiconductor
fiber is fabricated on the basis of silica glass doped by two
appropriate semiconductor compounds. In that case the nonlinear
correction to the medium's refractive index can be expressed in
the form $\delta n = n_{2}I-n_{4}I^{2}$, where $I$ being the light
intensity and the coefficients $n_{2} > 0$ and $n_{4} > 0$
determine the nonlinear response of the media. They are related to
third order susceptibility $\chi^{(3)}$ and fifth order
susceptibility $\chi^{(5)}$ through $n_{2} = 3\chi^{(3)}/8n_{0}$
and $n_{4} =-5\chi^{(5)}/32n_{0}$, where $n_{0}$ being the linear
refractive index. Although, formally it may be obtained by an
expansion of the saturable nonlinearity $\delta n = n_{2}I[1 +
(n_{4}/n_{2})I]^{-1}$, it is restricted under the effect of
self-focusing as $d(\delta n) = dI$.

The generalized CQNLSE with distributed nonlinear gain governing
the propagation of the optical field in a single-mode double-doped
optical fiber can be written in the form
 \bea
i\psi_{z}=\frac{\beta(z)}{2}\psi_{\tau\tau}-\gamma(z)\mid \psi
\mid^2 \psi+\delta(z)\mid \psi \mid^4 \psi+i g(z)\psi+i\chi(z)\mid
\psi \mid^2 \psi,\label{eq.1}
 \eea
where $\psi$  is the complex envelope of the electric field in a
comoving frame, $z$ is the propagation distance, $\tau$ is the
retarded time, $g(z)$ is the gain function, and $\chi(z)$ accounts
for the nonlinear gain or absorption \cite{gorza,li}. In the
absence of the last term and quintic nonlinear term, this equation
has exact similariton solutions that exhibit linear chirp. So
also, in the presence of nonlinear gain and in the absence of
quintic nonlinear term, this equation has exact chirped SSPs.
Recently, pedestal free compression of the pulses in the
generalized nonlinear Schr\"odinger equation with quintic
nonlinearity without nonlinear gain term has been reported
\cite{senthil}. Also, chirped and chirp free self-similar cnoidal
and solitary wave solutions of this model without nonlinear gain
term have been analyzed~\cite{dai}. But here we are concerned with
similariton pulses characterized by a nonlinear chirp, resulting
from the nonlinear gain. For finding the similariton solutions
using the FT of Eq. (1), one writes the complex function
$\psi(z,\tau)$ as \be \psi(z,\tau)=P(z,\tau){\rm exp}\{im_{0}{\rm
ln}[P(z,\tau)]+i\Phi(z,\tau)\}
 \ee
where $m_{0}$ denotes the nonlinear chirp parameter, and P and
$\Phi$ are real functions of $z$ and $\tau$. Using this ansatz, we
find the system of two equations for phase $\Phi(z,\tau)$, and
amplitude $P(z,\tau)$ \bea
P\Phi_{z}=\frac{\beta(z)}{2}\left[2m_{0}P_{\tau}\Phi_{\tau}+P\Phi^{2}_{\tau}+
\frac{m_{0}P^{2}_{\tau}}{P}-P_{\tau\tau}\right]+
\gamma(z)P^{3}-\delta(z)P^{5}-m_{0}P_{z}, \eea \bea
P_{z}=\frac{\beta(z)}{2}\left[2P_{\tau}\Phi_{\tau}+P\Phi_{\tau\tau}+
\frac{m_{0}P^{2}_{\tau}}{P}+m_{0}P_{\tau\tau}\right]+
\g(z)P+\chi(z)P^{3}. \eea In general case when the coefficients of
the CQNLSE with gain and nonlinear gain are the functions of the
distance $z$, the amplitude $P(z,\tau)$ of the similariton
solutions has the form \be P(z,\tau)=\frac{Q(T)}{\sqrt{\Gamma(z)}}
{\rm exp}(S(z)-m_{0}\Theta(z)),\ee where the scaling variable $T$
is given by \be T=\frac{\tau-\tau_{p}(z)}{\Gamma(z)}.  \ee And the
other functions $R(z)$, $\Theta(z)$, $S(z)$ and $\tau_{p}(z)$ take
the forms

\be
  R(z)=\int_{0}^{z}\beta(z^\prime)dz^\prime,
  \ee

\be
\Theta(z)=-\lambda\int_{0}^{z}\frac{\beta(z^\prime)}{[1-2c_{0}R(z^\prime)]^2}dz^\prime,
  \ee

\be
  S(z)=\frac{1}{2}{\rm ln}\left(\frac{m^{2}_{0}-2}{2}
  \rho(0)\right)+\int_{0}^{z}g(z^\prime)dz^\prime,\label{eq.8}
  \ee
\be \tau_{p}(z)=\tau_{c}-b_{0}R(z) ,\label{eq.9}\ee
 where $b_{0}$, $\lambda$, $c_{0}$ and $\tau_{c}$ are the integration
  constants and  $\rho(z)=\b(z)/\g(z)$.
  Here $\Gamma(z)$ and $Q(T)$ are some functions we seek, where
  without loss of generality we can assume $\Gamma(0)=1$. Let us
  assume that the phase has the quadratic form
  \be
\Phi(z,\tau)=a(z)+b(z)[\tau-\tau_{p}(z)]+c(z)(\tau-\tau_{p})^2,
\label{eq.3}\ee where the pulse position $\tau_{p}$ is a function
of $z$. Thus we consider the class of similariton solutions with
the phase given by Eq. (11). Then Eq. (3) with the phase (11) can
be written \bea
PM=\frac{\beta(z)}{2}N+\gamma(z)P^{3}-\delta(z)P^{5}+
m_{0}\frac{(d\Gamma/dz)}{2\Gamma}P-m_{0}\left[\frac{dS}{dz}-m_{0}\frac{d\Theta}{dz}\right]P
\eea where \bea
M=\frac{da}{dz}+\left(\frac{db}{dz}-2c(z)\frac{d\tau_{p}(z)}{dz}\right)(\tau-\tau_{p}(z))-b(z)\frac{d\tau_{p}(z)}{dz}+
\frac{dc(z)}{dz}(\tau-\tau_{p}(z))^{2},~~~~~~\eea \bea
N=P\left[b(z)+2c(z)(\tau-\tau_{p}(z))\right]^{2}+2m_{0}P_{\tau}[b(z)+2c(z)(\tau-\tau_{p}(z))]
+\frac{m_{0}P^{2}_{\tau}}{P}-P_{\tau\tau}.~~~~~~~~\eea Equation
(12) contains an explicit dependence on the variable
$(\tau-\tau_{p})$ which disappears when the terms at monomial
$(\tau-\tau_{p})^2$ are equals, hence we find three equations \be
\frac{dc(z)}{dz}=2\beta(z)c(z)^{2} \ee \bea
-P\frac{db(z)}{dz}+2Pc(z)\frac{d\tau_{p}}{dz}=\frac{\beta(z)}{2}
\left[-4m_{0}P_{\tau}c(z)-4Pb(z)c(z)\right], \eea
\bea
-P\frac{da(z)}{dz}+2Pb(z)\frac{d\tau_{p}}{dz}=\frac{\beta(z)}{2}U-\gamma(z)P^{3}+\delta(z)P^{5}-
m_{0}\frac{(d\Gamma/dz)}{2\Gamma}P+m_{0}\left[\frac{dS}{dz}-m_{0}\frac{d\Theta}{dz}\right]P
 ~~~~~~~~~\eea
 where
 \be
 U=P_{\tau\tau}-m_{0}\frac{P^{2}_{\tau}}{P}-2m_{0}b(z)P_{\tau}-Pb(z)^{2}.
 \ee
 Combining Eqs. (4) and (11) we find a third equation for our
 general class of similariton autonomous solutions with quadratic
 phase
 \bea
 P_{z}=\beta(z)c(z)P+\beta(z)b(z)P_{\tau}+2\beta(z)c(z)(\tau-\tau_{p}(z))P_{\tau}+
\frac{\beta(z)}{2}\left[m_{0}P_{\tau\tau}+m_{0}\frac{P^{2}_{\tau}}{P}\right]+g(z)P+\chi(z)P^{3}.
 ~~~~~\eea
 Taking into account Eq. (5) we find that Eq. (19) will be satisfied
 if and only if the function $\Gamma(z)$ is defined as
 \be
\frac{1}{\Gamma(z)}\frac{d\Gamma(z)}{dz}=-2\beta(z)c(z).
 \ee
The solutions of Eqs. (15), (16), and (20) are \be
c(z)=\frac{c_{0}}{1-2c_{0}R(z)}, \ee \be
P_{\tau}=\frac{b_{0}}{m_{0}\beta}\frac{dR}{dz}P-\frac{b_{0}P}{m_{0}},
\ee where $b(z)=b_{0}$, and \be \Gamma(z)=1-2c_{0}R(z). \ee Taking
into account Eqs. (5) and (17) we find \bea
\frac{d^{2}Q}{dT^{2}}-VQ -\frac{2\gamma\Gamma}{\beta}{\rm
exp}[2S-2m_{0}\Theta]Q^{3}+
 \frac{2\delta}{\beta}{\rm
exp}[4S-4m_{0}\Theta]Q^{5}=0.
 \eea
 where
 \bea
 V=\Gamma^{2}\frac{b^{2}_{0}}{m_{0}\beta^{2}}(dR/dz)^{2}-
\Gamma^{2}\frac{2b^{2}_{0}}{m_{0}\beta}(dR/dz)+\Gamma^{2}\frac{b^{2}_{0}}{m_{0}}-
\Gamma^{2}b^{2}_{0} -\Gamma^{2}\frac{2}{\beta}(da/dz)-\nonumber\\
\frac{m_{0}\Gamma}{\beta}\frac{d\Gamma}{dz}-\frac{2m_{0}\Gamma^{2}}{\beta}
(\frac{dS}{dz}-m_{0}\frac{d\Theta}{dz}).
 \eea
 In general case, the coefficients in Eq. (24) are functions of
 variable $z$ but the function $Q(T)$ depends only on the scaling
 variable $T$, hence this equation has nontrivial solutions [$Q(T)\neq
 0$] if and only if the coefficients in Eq. (24) are constants
 \bea
 \lambda=\Gamma^{2}\frac{b^{2}_{0}}{m_{0}\beta^{2}}(dR/dz)^{2}-
\Gamma^{2}\frac{2b^{2}_{0}}{m_{0}\beta}(dR/dz)+\Gamma^{2}\frac{b^{2}_{0}}{m_{0}}-
\Gamma^{2}b^{2}_{0} -\Gamma^{2}\frac{2}{\beta}(da/dz)-\nonumber\\
\frac{m_{0}\Gamma}{\beta}\frac{d\Gamma}{dz}-\frac{2m_{0}\Gamma^{2}}{\beta}
(\frac{dS}{dz}-m_{0}\frac{d\Theta}{dz}),
 \eea
 \be
 \frac{\gamma\Gamma}{\beta}{\rm exp}[2S-2m_{0}\Theta]=\alpha,
 \ee
 \be
\frac{\delta}{\beta}{\rm exp}[4S-4m_{0}\Theta]=\epsilon.
 \ee
 Here $\lambda=const$, $\alpha=const$, hence Eqs. (26) and (27)
 yield
 \bea
 \lambda=b^{2}_{0}(\frac{1-m_{0}}{m_{0}})-\frac{2}{\beta(0)}\frac{da}{dz}\mid_{z=0};~~~
 \alpha=\frac{m^{2}_{0}-2}{2}
 \eea
because $\Gamma(0)=1$ and $S(0)=\frac{1}{2}{\rm
ln}\left(\frac{m^{2}_{0}-2}{2}
  \rho(0)\right)$. Thus in nontrivial case Eq.
(24) can be written as \be \frac{d^{2}Q}{dT^{2}}-\lambda Q-2\alpha
Q^{3}+2\epsilon Q^{5}=0 \ee The solution of Eq. (26) is \be
a(z)=a_{0}+\frac{1+m^{2}_{0}}{2}\Theta(z)-\frac{b^{2}_{0}}{2}R(z)-\frac{m_{0}}{2}{\rm
ln}[\mid c(z)\mid]-m_{0}S(z) \ee It is useful to write Eq. (27) in
the form \be \rho(z)=\rho(0)[1-2c_{0}R(z)]{\rm
exp}[2S-2m_{0}\Theta] \ee where we define the function $\rho(z)$
as \bea \rho(z)=\frac{\beta(z)}{\gamma(z)};~~~
\rho(0)=\frac{\beta(0)}{\gamma(0)}=\frac{1}{\alpha}. \eea One may
differentiate this equation and find \be
g(z)=\frac{1}{2\r(z)}\frac{d}{dz}\r(z) + \frac{c_0\b(z)}{1-2c_0
R(z)}-\frac{\mu\l m_0\b(z)}{\left[1-2c_0
R(z)\right]^2},\label{eq.13}
  \ee

  \be
\frac{\chi(z)}{\g(z)}=\frac{3m_{0}}{m^{2}_{0}-2}\label{eq.14}
  \ee
  where $\m=1$ or $\frac{3}{4}$. The former condition describes that the
parameter functions in Eq. (1) can not be chosen independently,
while the latter condition implies that the nonlinear chirp
parameter $m_{0}$ in fact be determined by the ratio
$\chi(z)/\g(z)$. In this paper, it requires that this ratio is a
constant. From the physical cases we have considered, we come to
the conclusion that $m^{2}_{0}\neq 2$ for arbitrary nonlinear
materials.
\section{Exact similariton solutions}
  In order to obtain the exact similariton solutions of Eq. (1), we use a fractional transform
\be Q(T)=\frac{A+B{f}^2(T)}{1+D{f}^2(T)},
  \ee
that connects Eq. (30) to the elliptic equation:
$f^{\prime\prime}\pm af\pm b f^{3}=0$, where $a$ and $b$ are real.
As is well known, $f$ can be taken as any of the three Jacobian
elliptic functions with an appropriate modulus parameter:
cn$(T,m)$, dn$(T,m)$, and sn$(T,m)$, with amplitude and width,
appropriately depending on $m$. Using the limiting conditions of
the modulus parameter, one can obtain both localized and
trigonometric solutions. We list below a few localized as well as
periodic solitary wave solutions.

For explicitness, we consider Eq. (30), with all the parameters
and illustrate below various types of solutions of Eq. (1), by
taking $f=cn(T,m)$. Other cases can be deduced by using the
Jacobian elliptic function identities: ${\rm dn}^{2}(T,m)=1-m{\rm
sn}^{2}(T,m)$ and ${\rm sn}^{2}(T,m)=1-{\rm cn}^{2}(T,m)$. The
consistency conditions are given by,
 \bea
-A\lambda+2(AD-B)(m-1)-2\a A^{3}+2\epsilon A^{5}=0,~~~~~\\
 2(AD-B)[2(1-2m)+3D(1-m)]+4D(AD-B)(m-1)-\lambda(2AD+B)-\nonumber\\
 2\a(3A^{2}B+2A^{3}D)+10\epsilon A^{4}B=0,~~~~~\\
2(AD-B)[2D(2m-1)+3m]+4D(AD-B)[2(1-2m)+3D(1-m)]+2D^{2}(AD-B)(m-1)-\nonumber\\
\lambda(6AD^{2}+2BD)-2\a (3AB^{2}+6A^{2}BD+A^{3}D^{2})+20\epsilon
A^{3}B^{2}=0,~~~~~~\\
-2mD(AD-B)+4D(AD-B)[2D(2m-1)+3m]+2D^{2}(AD-B)[2(1-2m)+3D(1-m)]-\nonumber\\
\lambda(4AD^{3}+6BD^{2})-2\a(B^{3}+6AB^{2}D+3A^{2}BD^{2})+20\epsilon
A^{2}B^{3}=0,~~~~~~~\\
-4mD^{2}(AD-B)+2D^{2}(AD-B)[2D(2m-1)+3m]-\lambda(AD^{4}+BD^{3})-\nonumber\\2\a
(2B^{3}D+3AB^{2}D^{2})+10\epsilon AB^{4}=0,~~~~~~~~\\
-2mD^{3}(AD-B)-\lambda BD^{4}-2\a B^{3}D^{2}+2\epsilon
B^{5}=0.~~~~~~~
 \eea
 The above equations clearly indicate that the solutions, for
$m=1$, $m=0$ and other values of $m$, have distinct properties. To
find the solution for either $m=1$ or $m=0$ or other values of $m$
we should keep all the parameters in the FT. Here, we list three
such cases: $m=0$, $m=1$, and $m=1/2$ in detail.
\subsection{Trigonometric solution}
  For $m=0$, from Eq. (42) we find that
  \be
  B=\sqrt{\Gamma^{\prime}}D
  \ee
  where $$
  \Gamma^{\prime}=\frac{\a\pm\sqrt{\a^{2}+2\epsilon\lambda}}{\epsilon}.
  $$
  From Eq. (41) we find that
  \be
  A=\frac{[4(1-\lambda)-4\a\Gamma^{\prime}]\sqrt{\Gamma^{\prime}}}{4-(10\epsilon{\Gamma^{\prime}}^{2}-6\a\Gamma^{\prime}-\lambda)}
  \ee
  And we find that
  \be
  D=\frac{(4+2\lambda)\sqrt{\Gamma^{\prime}}-(-2\a+20\epsilon\Gamma^{\prime})A^{3}+12\a A^{2}\sqrt{\Gamma^{\prime}}-(-6\lambda-6\a\Gamma^{\prime}+4)A}
  {10A-10\sqrt{\Gamma^{\prime}}}
  \ee
  Hence, the trigonometric solution of Eq. (1) can be written as
\bea \psi(z,\tau)=\frac{\sqrt{(m^{2}_{0}-2)(\b(z)/\g(z))}}
{\sqrt{2}\left[1-2c_{0}R(z)\right]}\frac{A+B{\rm
cos}^{2}\left[\frac{\tau-\tau_{c}+b_{0}R(z)}{1-2c_{0}R(z)}\right]}{1+D{\rm
cos}^{2}\left[\frac{\tau-\tau_{c}+b_{0}R(z)}{1-2c_{0}R(z)}\right]}\nonumber\\
\times~~{\rm exp}\left\{im_{0}{\rm ln}\left[\frac{A+B{\rm
cos}^{2}\left[\frac{\tau-\tau_{c}+b_{0}R(z)}{1-2c_{0}R(z)}\right]}{1+D{\rm
cos}^{2}\left[\frac{\tau-\tau_{c}+b_{0}R(z)}{1-2c_{0}R(z)}\right]}\right]+i\Phi(z,\tau)\right\}.~~\label{eq.21}
\eea
 We emphasize that this trigonometric
solution is the general solution of this model and is valid for
all values of the amplitude parameters A, B, and D subject to the
condition $AD-B\neq 0$. This solution of Eq. (1) has no analogue
in the limit $n_{4}\rightarrow 0$.
\subsection{Hyperbolic solution}
For $m=1$, from Eq. (37) we are able to determine the amplitude
parameter $A$ completely as \be A^{2}=
\frac{\a\pm\sqrt{\a^{2}+2\epsilon\lambda}}{2\epsilon}.\ee From Eq.
(38) we find that $B$ and $D$ are related to each other as \be
B=\Gamma_{2}D, \ee where $$ \Gamma_{2}=\frac{A(4\a
A^{2}+2\l+4)}{10\e A^{4}-6\a A^{2}-\l+4}.
$$
Using this value of $B$ in Eqs. (39) and (40) we fully determine
$D$ as \be
D=\frac{10A-6\G_{2}}{4\G_{2}[1+\a(A+\G_{2})]-8A-\G_{2}-\frac{2\l\G^{2}_{2}}{A-\G_{2}}}.
\ee Thus, the solitary wave solution of Eq. (1) is written as

\bea \psi(z,\tau)=\frac{\sqrt{(m^{2}_{0}-2)(\b(z)/\g(z))}}
{\sqrt{2}\left[1-2c_{0}R(z)\right]}\frac{A+B{\rm
sech}^{2}\left[\frac{\tau-\tau_{c}+b_{0}R(z)}{1-2c_{0}R(z)}\right]}{1+D{\rm
sech}^{2}\left[\frac{\tau-\tau_{c}+b_{0}R(z)}{1-2c_{0}R(z)}\right]}\nonumber\\
\times~~{\rm exp}\left\{im_{0}{\rm ln}\left[\frac{A+B{\rm
sech}^{2}\left[\frac{\tau-\tau_{c}+b_{0}R(z)}{1-2c_{0}R(z)}\right]}{1+D{\rm
sech}^{2}\left[\frac{\tau-\tau_{c}+b_{0}R(z)}{1-2c_{0}R(z)}\right]}\right]+i\Phi(z,\tau)\right\}.~~
\eea
Again, we emphasize that this solution of Eq. (1) has no analogue
in the limit $n_{4}\rightarrow 0$.
\subsection{Pure cnoidal solution}
In order to obtain pure cnoidal wave solution, we put $m=1/2$ and
$B=D$ in the consistency conditions. From Eq. (40) we determine
the amplitude parameter $A$ completely as \be A=
\frac{-b\pm\sqrt{b^{2}-4ac}}{2a},\ee where $a=20\e-6\a$,
$b=3-4\l-12\a$, and $c=10(\e-\a)-11\l-2\a-3$. We determine the
value of $D$ by solving Eq. (42) as \be D=\frac{A-1}{2(\e-\a)-\l}.
\ee Hence, the pure cnoidal wave solution of Eq. (1) with all the
amplitude parameters is given by
\bea \psi(z,\tau)=\frac{\sqrt{(m^{2}_{0}-2)(\b(z)/\g(z))}}
{\sqrt{2}\left[1-2c_{0}R(z)\right]}\frac{A+B{\rm
cn}^{2}\left[\frac{\tau-\tau_{c}+b_{0}R(z)}{1-2c_{0}R(z)},m\right]}{1+D{\rm
cn}^{2}\left[\frac{\tau-\tau_{c}+b_{0}R(z)}{1-2c_{0}R(z)},m\right]}\nonumber\\
\times~~{\rm exp}\left\{im_{0}{\rm ln}\left[\frac{A+B{\rm
cn}^{2}\left[\frac{\tau-\tau_{c}+b_{0}R(z)}{1-2c_{0}R(z)},m\right]}{1+D{\rm
cn}^{2}\left[\frac{\tau-\tau_{c}+b_{0}R(z)}{1-2c_{0}R(z)},m\right]}\right]+i\Phi(z,\tau)\right\}.~~
\eea
Once again, we emphasize that this solution of Eq. (1) has no
analogue in the limit $n_{4}\rightarrow 0$.

\begin{figure}
 \centering
 \includegraphics[height=2.0in,width=5.0in]{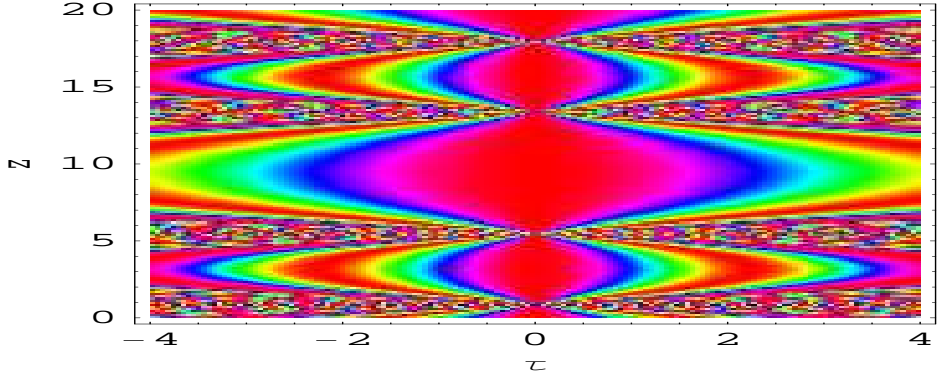}
 \caption{Density plot depicting the nonlinear chirping of the
trigonometric solution for $\sigma=0.5$, $\a = 3.45$, $\lambda =
-3.34$, and $\e=-2.78$}
\end{figure}

\section{Nonlinear chirping}
In this subsection, we wish to cite an example corresponding to
the trigonometric solution, illustrative of the fascinating
features of chirping in double-doped fibers, by considering the
system in which the GVD and the nonlinearity are distributed
according to \bea
  \beta(z)=\beta_0\cos(\s z),~~   \g(z)=\g_0
\cos(\s z),
  \eea
where $\beta_{0}$, $\g_{0}$ and $\s(\neq 0)$ are arbitrary
constants. In this case, the corresponding gain, and the nonlinear
gain of the fiber amplifier are given by \bea
g(z)=\frac{\s\nu\cos(\s z)}{2-2\nu\sin(\s z)}-\frac{\mu\l
m_{0}\cos(\s z)}{(1-\nu\sin(\s
z))^2}, \\
\chi(z)=\frac{3m_{0}\g_{0}}{m^{2}_{0}-2}\cos(\s z),
 \eea where the parameter $\nu=2c_{0} \beta_{0} /\s $ has been introduced
for brevity.

 Hence the amplitude of the similariton pulse is
\bea
  P(z,\tau)=\frac{\sqrt{(m^{2}_{0}-2)\rho(0)}}{\sqrt{2}(1-\nu{\rm sin}{\s}z)}\frac{A+B{\rm cos}^{2}
  \left[\frac{\tau-\tau_{p}(z)}{W(z)}\right]}{1+D{\rm cos}^{2}
  \left[\frac{\tau-\tau_{p}(z)}{W(z)}\right]} \eea

where
 $$ W(z)=\tau_{0}[1-\nu\sin(\s z)],$$ and the pulse position
$\tau_{p}$ varies with
$\tau_{p}=\tau_{c}-(b_{0}\beta_{0}/\s)\sin(\s z)$. The resultant
chirp consisting of linear and nonlinear contributions are derived
as \cite{Fermann} \bea
\delta\o(\tau)=\frac{m_{0}}{W(z)}\frac{\tan(\frac{\tau-
\tau_{p}}{W(z)})}{1+D\cos^{2}(\frac{\tau- \tau_{p}}{W(z)})}-
b_{0}-\frac{2c_{0}\tau_{0}}{W(z)}(\tau- \tau_{p}).~~ \label{eq.23}
\eea

We notice that the first term in Eq. (\ref{eq.23}) denotes the
nonlinear chirp that results from the nonlinear gain, while the
last two terms account for the linear chirp. The propagation of
this chirped pulse has been depicted in Fig. 1 for various
parameter values of $\tau_{0}$ and $\nu=1$.

\begin{figure}
 \centering
 \includegraphics[height=2.0in,width=5.0in]{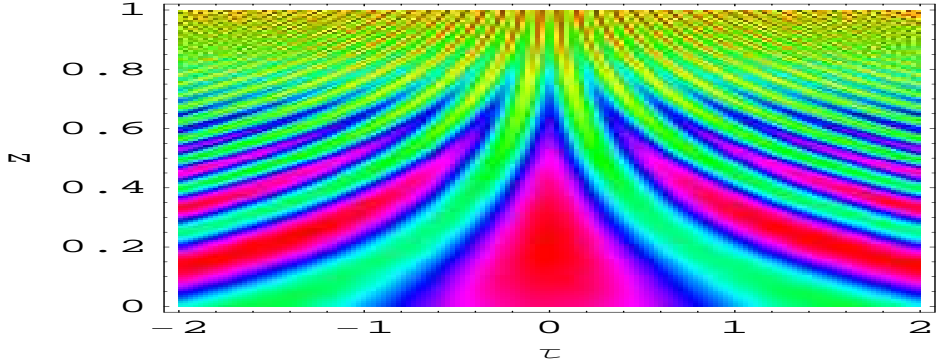}
 \caption{Density plot depicting the nonlinear compression of the
trigonometric solution for $\sigma=1.0$, $\a = 3.45$, $\lambda =
-3.34$, and $\e=-2.78$}
\end{figure}
\section{Nonlinear compression}
 Next, we elucidate the compression problem of the pulse in a
dispersion decreasing optical fiber. For the purpose of comparison
with Ref. \cite{kruglov,chen}, we assume that the GVD and the
nonlinearity are distributed according to the following relations
 \bea
  \beta(z)=\beta_0\exp(-\s z),~~  \g(z)=\g_0
\exp(\a z),
  \eea
where $\beta_0 \le 0$, $\g_0\ge 0$, and $\s\ne  0$. Then from Eq.
(34) the expression for gain can be calculated as \be
g(z)=-\a+\nu\s\frac{e^{-\s z}}{1-\nu(1-e^{-\s z})}-\frac{\mu\l
m_{0}\beta_{0}e^{-\s z}}{(1-\nu(1-e^{-\s z}))^2}.\label{eq.31}
  \ee

Let us consider the most typical physical situation when the loss
in an optical fiber is constant i.e., $\g(z)$ is constant.
According to Eq. (\ref{eq.31}), this occurs when $\nu =1$ and $\a
> 0$: $g(z)=-\a$, hence the gain is negative. It is remarkable
that the width of the solutions presented here tend to zero when
$z\rightarrow\infty$.

We apply the above insights to nonlinear compression of the
trigonometric solution given
  by Eq. (46). Then the amplitude of the compressed wave is
  \bea
  P(z,\tau)=U(z)\frac{A+B{\rm cos}^{2}
  \left[\frac{\tau-\tau_{p}(z)}{W(z)}\right]}{1+D{\rm cos}^{2}
  \left[\frac{\tau-\tau_{p}(z)}{W(z)}\right]}, \eea
  where $$
  U(z)=\frac{\sqrt{(m^{2}_{0}-2)\rho(0){\exp}{[-(\s+\a)z]}}}{\sqrt{2}[1-\nu(1-{\rm exp}{(-\s
  z)})]}.
  $$
 And
$W(z)=\tau_{0}[1-\nu(1-\exp(-\s z))].$ Fig.~2 shows that for
constant loss the trigonometric pulse can be compressed to any
required degree as $z\rightarrow\infty$, while maintaining its
respective original shape.
\section{Conclusion}
In conclusion, for generalized cubic-quintic nonlinear
Schr\"odinger-type equation with variable dispersion, variable
Kerr nonlinearity, variable gain or loss, and nonlinear gain, we
found exact chirped pulses that can propagate self-similarly
subject to simple scaling rules, of this model. The fact that the
pulse position of these chirped pulses can be precisely piloted by
appropriately tailoring the dispersion profile, is profitably
exploited to achieve optimal pulse compression of these newly
reported chirped similariton solutions. Studying soliton
bistability using the exact solutions found here will be an
interesting issue to be pursued. These analytical findings suggest
potential applications in areas such as optical fiber compressors,
optical fiber amplifiers, nonlinear optical switches, and optical
communications.
\section*{Acknowledgement}
This paper is dedicated to the fond memory of my father
Shri.~Thokala Ratna Raju, for his love and encouragement.


\section*{References}
\begin{thebibliography}{}

\bibitem{hasegawa} Hasegawa A and  Tappert F 1973 \emph{Appl.
      Phys. Lett.} {\bf 23}, 142
\bibitem{stolen} Mollenauer L F, Stolen R H and Gordon J P 1980 \emph{Phys. Rev. Lett.} {\bf  45}, 1095
 \bibitem{govind1} Agrawal G P  Nonlinear
  Fiber Optics, Academic Press, San Diego, 2001.
\bibitem{anderson} Anderson D, Desaix M, Karlsson M, Lisak M and
Quiroga-Teixeiro M L 1993 \emph{J. Opt. Soc. Am. B} {\bf 10}, 1185

\bibitem{Fermann} Fermann M E, Kruglov V I, Thomsen B C, Dudley J
M and Harvey J D 2000 \emph{Phys. Rev. Lett.} {\bf 84}, 6010 ;
Kruglov V I, Peacock A C, Dudley J M and Harvey J D 2000
\emph{Opt. Lett.} {\bf 25}, 1753.
\bibitem{kruglov1}
Kruglov V I, Peacock A C, Harvey J D and Dudley J M 2002 \emph{J.
Opt. Soc. Am. B} {\bf 19}, 461.
\bibitem{limpert} Limpert J et al 2002 \emph{Opt. Express} {\bf 10}, 628.
\bibitem{malinowski} Malinowski A et al 2004 \emph{Opt. Lett.} {\bf
29}, 2073.
\bibitem{peacock} Peacock A C, Kruhlak R J, Harvey J D and Dudley J M 2002 \emph{Opt. Commun.} {\bf 206}, 171.
\bibitem{finot}Finot C, Millot G, Pitois S, Billet C and Dudley J M 2004 \emph{IEEE J. Sel. Top. Quantum Electron.} {\bf 10}, 1211.
\bibitem{Ilday} Ilday F \"{O}, Buckley J R, Clark W G and Wise F W 2004 \emph{Phys. Rev.
Lett.} {\bf 92}, 213902.
\bibitem{moores}Moores J D 1996 \emph{Opt.
Lett.} {\bf 21}, 555.
\bibitem{serkin} Serkin V N and Hasegawa A 2000 \emph{Phys. Rev. Lett.} {\bf 85}, 4502.
\bibitem{kruglov}Kruglov V I, Peacock A C and Harvey J D 2003 \emph{Phys. Rev.
Lett.} {\bf 90}, 113902.
\bibitem{chen} Chen S and Yi L 2005 \emph{Phys. Rev. E} {\bf 71}, 016606.
\bibitem{chen1}Chen S, Yi L, Guo D S and Lu P 2005 \emph{Phys. Rev. E} {\bf  71}, 016622.
\bibitem{mechin}M\'echin D, Im S H, Kruglov V I, Harvey J D 2006 \emph{Opt. Lett.} {\bf 31}, 3106.
\bibitem{angelis} Angelis C D 1994 \emph{IEEE J.
Quant. Electron.} {\bf 30}, 818.

\bibitem{abdullaev}Abdullaev F K, Gammal A, Tomio L and Frederico T 2001 \emph{Phys.
Rev. A} {\bf 63}, 043604. \bibitem{xhang} Xhang W, Wright E M, Pu
H and Meystre P 2003 \emph{Phys. Rev. A} {\bf 68}, 023605.
\bibitem{inouye} Inouye S et al 1998 \emph{Nature (London)} {\bf 392}, 151.
\bibitem{serkin2} Serkin V N, Chapela V M, Percino J and Belyaeva T L 2001 \emph{Opt. Commun.} {\bf 192}, 237.
\bibitem{ndzana} Ndzana F I, Mohamadou A and Kofan\'e T C 2007 \emph{Opt. Commun.} {\bf 275}, 421.
\bibitem{avelar}Avelar A T, Bazeia D Cardoso W B 2009 \emph{Phys.
Rev. E} {\bf 79}, 025602.
\bibitem{adib} Adib B, Heidari A and Tayyari S F 2009 \emph{Commun. Nonlinear Sci. Numer.
Simulat.} {\bf 14}, 2034.
\bibitem{triki} Azzouzi F et al 2008 \emph{Chaos, Solitons and
Fractals} {\bf 39}, 1304.

\bibitem{serkin3}Porsezian K et al 2009 \emph{IEEE J. Quntum
Electronics} {\bf 45}, 1577.

\bibitem{serkin4} Serkin V N, Hasegawa A and Belyaeva T L 2010
\emph{Phys. Rev. A} {\bf 81}, 023610.

\bibitem{panigrahi} Soloman Raju T, Nagaraja Kumar C and Panigrahi P K 2005 \emph{J. Phys. A: Math. and Gen.} {\bf 38},  L271.
\bibitem{gorza}Gorza S P, Roig N, Emplit Ph and Haelterman M 2004
\emph{Phys. Rev. Lett.} {\bf 92}, 084101 .
\bibitem{li}Li Z et al 2002 \emph{Phys. Rev. Lett.} {\bf 89}, 263901 .
\bibitem{senthil} Senthilnathan K, Li Q, Nakkeeran K and Wai P K A 2008 \emph{Phys. Rev.
A} {\bf 78}, 033835.
\bibitem{dai} Dai C, Wang Y and Yan C 2010 \emph{Opt. Commun.} {\bf 283}, 1489.
\end{thebibliography}
\end{document}